\documentclass[superscriptaddress,twocolumn,amsmath,amssymb,nofootinbib,longbibliography]{revtex4-1}
\usepackage[normalem]{ulem}

\usepackage{color}
\usepackage[usenames,dvipsnames]{xcolor}

\definecolor{bleudefrance}{rgb}{0.19, 0.55, 0.91}

\newcommand{\kt}{K}                                            

\newcommand{\be}{\begin{equation}}             
\newcommand{\ee}{\end{equation}}               
\newcommand{\ba}{\begin{eqnarray}}             
\newcommand{\ea}{\end{eqnarray}}               

\begin{document}

\title{Homogeneous Symmetry Operators in Kerr--NUT--AdS Spacetimes}

\author{Finnian Gray}
\email{finnian.gray@univie.ac.at}
\affiliation{University of Vienna, Faculty of Physics, Boltzmanngasse 5, A 1090 Vienna, Austria
}
\author{David Kubiz\v{n}\'{a}k}
\email{david.kubiznak@matfyz.cuni.cz}
\affiliation{Institute of Theoretical Physics, Faculty of Mathematics and Physics,
Charles University, Prague, V Hole\v{s}ovi\v{c}k\'ach 2, 180 00 Prague 8, Czech Republic}

\date{April 12, 2024}         

\begin{abstract}
It is well known that the Kerr--NUT--AdS spacetimes possess hidden symmetries encoded in the so-called principal Killing--Yano tensor. In this paper, focusing on the four-dimensional case, we obtain a number of symmetry operators for scalar, vector, and tensor perturbations, that are of degree two (to be defined below) and homogeneous in the principal tensor. In particular, by considering homogeneous operators that are linear, quadratic, and cubic in the principal tensor, we recover a  complete set of 4 mutually commuting operators for scalar perturbations, underlying the separability of (massive) scalar wave equation. Proceeding to vector and tensor perturbations of the Kerr--NUT--AdS spacetimes, we find a set of 7 and 8 commuting operators, respectively. It remains to be seen whether such operators can be used to separate the corresponding spin 1 and spin 2 test field equations in these spacetimes.     
\end{abstract}

\maketitle

\section{Introduction}

Hidden symmetries play an important role in black hole physics. 
As opposed to explicit symmetries, that are `visible' in the spacetime and are encoded in Killing vector fields, hidden symmetries correspond to genuine symmetries of the phase space, e.g. \cite{crampin1984hidden, Cariglia:2014ysa}, and are described by higher rank tensors that obey a natural generalization of the Killing vector equation. Among these, perhaps the most remarkable is the hidden symmetry of the {\em Principal Killing--Yano (PKY) tensor}, see \cite{Frolov:2017kze} for a review.

The PKY tensor is a non-degenerate closed conformal Killing--Yano 2-form $k_{ab}$. This is an object whose covariant derivative is completely encoded in its divergence
\be\label{xi}
\xi_a=\frac{1}{d-1}\nabla^b k_{ba}\,,
\ee
{where $d$ stands for the number of spacetime dimensions.}
This divergence is a Killing vector by construction and explicitly the PKY tensor satisfies the following defining equation:
\be\label{PKYT}
\nabla_c k_{ab}=g_{ca}\xi_b-g_{cb}\xi_a\,.
\ee

The principal tensor {exists in the} Kerr spacetime and  its `square' gives rise to  Carter's famous constant for geodesics \cite{Carter:1968rr}. More generally, the PKY tensor can be found in a large class of the (off-shell) Kerr--NUT--AdS spacetimes in any number of dimensions \cite{Kubiznak:2006kt, Krtous:2008tb, Houri:2008ng}, and underlies many of its remarkable properties, such as complete integrability of the geodesic equations \cite{Page:2006ka}, separability of the Hamilton--Jacobi and Klein--Gordon equations \cite{Frolov:2006pe}, separability of the Dirac equation \cite{Oota:2007vx}, or the special algebraic type of these metrics \cite{Mason:2010zzc}.

More recently \cite{Lunin:2017drx,Frolov:2018ezx}, it was also shown how, using the PKY tensor, one can construct 
a `polarization tensor' $B_{ab}=(g_{ab}+\lambda k_{ab})^{-1}$, which gives rise to a separable ansatz for the (massive) vector perturbations:
\be\label{eq: LFKK ansatz}
A^a=B^{ab}\partial_b Z\,, 
\ee
where $Z$ is the scalar function that takes a standard separable form, and $\lambda$ is the additional separation constant, see  \cite{Frolov:2018ezx} for details.

Interestingly, in the scalar \cite{Carter:1977pq, Sergyeyev:2007gf}, conformal scalar~\cite{Gray:2020rtr,Gray:2021wzf}, and the Dirac \cite{Carter:1979fe, Cariglia:2011qb} cases, the separation can be linked to the existence of a {\em compete set} of the mutually {\em commuting operators} (called symmetry operators) whose common eigenfunction is the separable solution. Remarkably, such operators take a simple form that is homogeneous (linear, quadratic, cubic, and so on) in the principal tensor and contains a fixed number of derivatives (some of which may act on the principal tensor while others are free act on the field). We call this number the {\em degree} of the corresponding operator. In this work, we study such {\em homogeneous operators} and restrict ourselves to degree two. Such operators can be of the first-order or second-order in the number of derivatives acting on the field.

For example, as we shall review in Sec.~\ref{Sec:scalar} the separability of the massive scalar equation is characterized by the existence of 3 homogeneous {degree two  operators that are linear,} quadratic, and cubic in the PKY tensor. 
Together with the scalar wave operator itself, these form a {\em complete set} of mutually commuting operators that underlie the separability of the scalar field equations. As discussed in App.~\ref{App:Dirac}, similar result remains true also for the Dirac case. 

{For vector and tensor perturbations, however, the situation is less straightforward. While such perturbations can be tackled with the help of the Teukolsky equation \cite{Teukolsky:1973ha, Teukolsky:1972my} (see \cite{Dolan:2021ijg,Dolan:2023enf} and the references therein for recent developments on metric perturbations in the Teukolsky formalism), the relationship to the principal tensor and whether there exists an underlying set of mutually commuting operators is much less clear in this case. One must employ various reconstruction procedures to obtain the fields from the Teukolsky scalars~\cite{Chrzanowski:1975,Kegeles:1979,Wald:1979PRL} which is an ongoing field of study~\cite{Green:2019nam,Toomani:2021jlo,Green:2022htq}. See \cite{Andersson:2014lca,Aksteiner:2016pjt,Aksteiner:2016mol,Michishita:2019eeh,Tsuchiya:2020hur,Jacobsson:2022mls} for related works on symmetry operators for various test field equations. Moreover, contemporary work \cite{Houri:2019lnu} constructs the required symmetry operators for the Maxwell case, using the equations resulting from the ansatz \eqref{eq: LFKK ansatz}. 
{However, recently there has been a new breakthrough achieved in \cite{Mei:2023pho}, where some homogeneous operators were constructed directly for the field equations of the vector and tensor perturbations of the vacuum Kerr spacetime.}

It is the aim of this paper, to systematically study such homogeneous symmetry operators, picking up the threads of this recent work \cite{Mei:2023pho}. In this way we discover novel symmetry operators for the vector and tensor perturbations of the Kerr--NUT--AdS spacetimes. 
Notably, contrasting with the recent work, our basic building block is directly the principal tensor $k$, instead of its dual Killing--Yano tensor $f=*k$. Since the PKY tensor comes to the fore for Kerr--NUT--AdS spacetimes in arbitrary number of dimensions, our results are readily generalizable to $d$ dimensions (with possibly $d$-dependent) factors, thus providing symmetry operators for black hole perturbations in higher dimensions \cite{myprep:2024}.

Our work is organized as follows. In the next section, focusing on four dimensions, we overview the (off-shell) Kerr--NUT--AdS spacetimes and their remarkable symmetries. In Sec.~\ref{Sec:scalar} we review the homogeneous symmetry operators for the (massive) scalar field, their systematic derivation, and application to separability of the scalar field equation. In Sec.~\ref{Sec:vector} we systematically study symmetry operators for the (massive) vector perturbations, while Sec.~\ref{Sec:tensor} is devoted to the massive tensor perturbations.  
We conclude in Sec.~\ref{Sec:summary}.
In App.~\ref{App:Dirac} we review the symmetry operators for the Dirac equation and its separability. App.~\ref{App:motion} contains analogous study of constants of motion for geodesic and classical spinning particle trajectories. {App.~\ref{App:f} explores the possibility of using the Hodge dual of the PKY tensor in the study of linear vector symmetry operators.      
}

\section{Kerr--NUT--AdS and its hidden symmetries}\label{Sec:KerrNUTAdS}
In this paper we shall study various perturbations of four-dimensional Kerr--NUT--AdS spacetimes, which we write in a convenient Carter-like form \cite{Carter:1968cmp}:
\ba
g&=&-\frac{\Delta_r}{\Sigma}(d\tau+y^2 d\psi)^2+\frac{\Delta_y}{\Sigma}(d\tau-r^2 d\psi)^2\nonumber\\
       &&+\frac{\Sigma}{\Delta_r}dr^2+\frac{\Sigma}{\Delta_y}dy^2\,, 
\ea
where 
\be\label{Sigmadef}
\Sigma=\sqrt{-\det (g) }=r^2+y^2\,,
\ee
and the metric functions $\Delta_r$ and $\Delta_y$ are functions of one variable:
\be \label{generalDelta}
\Delta_y=\Delta_y(y)\,,\quad \Delta_r=\Delta_r(r)\,.
\ee
In what follows we shall consider three cases. i) The {\em off-shell canonical} metric, for which the metric functions take the above general form \eqref{generalDelta}. 
Many of the geometric properties remain valid in this general case \cite{Frolov:2017kze}.  
ii) The {\em Einstein space} case, 
when the metric solve the vacuum Einstein equations with the cosmological constant $\Lambda$, and the corresponding 
metric functions take the following specific form: 
\be\begin{split}\label{Delr}
\Delta_r&=(r^2+a^2)(1-\Lambda r^2/3)-2Mr\, ,\\
\Delta_y&=(a^2-y^2)(1+\Lambda y^2/3)+2Ny\, .
\end{split}
\ee
Here, $M$ stands for the mass parameter, $a$ denotes the rotation, and $N$ is the  Newmann--Unti--Tamburino (NUT)~\cite{NewmanEtal:1963} parameter. iii) Finally, we shall consider the (pure) {\em vacuum case}, for which the metric functions take the specific form \eqref{Delr}, with  
\be
\Lambda=0\,. 
\ee
The metric then becomes the vacuum solution of the Einstein equations -- the standard Kerr spacetime \cite{Kerr:1963ud} equipped with the NUT parameter $N$.

The remarkable property of the off-shell canonical metric (inherited by its special on-shell subcases) is that it admits the powerful symmetry of the PKY tensor $k$, obeying \eqref{PKYT}. 
Explicitly, it is given by
\be
k= y dy\wedge(d\tau-r^2 d\psi)
-r dr\wedge (d\tau+y^2 d\psi)\,.
\ee
This tensor generates all of the explicit and hidden symmetries, and determines many remarkable properties of the geometry, many of which prevail for higher-dimensional Kerr--NUT--AdS spacetimes as well, see \cite{Frolov:2017kze}.

In particular, it is possible to show that $\xi$, from \eqref{xi}, is given by 
\be 
\xi=\partial_\tau\,.
\ee
{We call it the primary Killing vector. Since, due to the Killing vector equation, its covariant derivative is completely antisymmetric, the complete information about two derivatives of the PKY tensor $k$ is encoded in $(d \xi)$. In other words, the following 3 objects encode information about $k$, and its first, and second order derivatives:
\be\label{k2der}
k_{ab}\,,\quad \xi^a\,,\quad (d\xi)_{ab}\,. 
\ee
In the following sections, 
such objects will be used to construct symmetry operators for various test field equations. 
}
 
{The `square' of the principal tensor gives rise to a Killing tensor
\be\label{CKT and KT explicit expressions}  
 \kt_{ab}=(k^2)_{ab}+\frac12 k^2 g_{ab}\,,
 \ee
 obeying $\nabla_{(c}K_{ab)}=0$. Here, we have introduced   
\ba
(k^2)_{ab}&=&k_{ac}k^c{}_b\,,\quad 
 k^2=k_{ab}k^{ab}\,,\nonumber\\
\ea
and denote for future convenience 
\ba
(k^3)_{ab}&=&k_{ac}(k^2)^c{}_b\,,\quad k^3=(k^3)^a{}_a=0\,. 
\ea 
Moreover, the following vector, cubic in the PKY tensor:
\be \label{zeta}
\zeta^a=K^a{}_b \xi^b=(\partial_\psi)^a\,,
\ee
is the secondary Killing vector of the spacetime. 
} 

Finally, the Hodge dual of the PKY tensor is a Killing--Yano tensor \cite{Frolov:2017kze}
\be 
f=*k=rdy\wedge(d\tau-r^2d\psi)+ydr\wedge (d\tau+y^2d\psi)\,,
\ee
a generalization of the Killing vector, obeying $\nabla_{(c}f_{a)b}=0$.
It is related to the above Killing tensor as follows: $K_{ab}=f_{ac}f^c{}_b$. Since its covariant derivative is completely antisymmetric, it is encoded in $(df)$. Moreover, due to the relation \eqref{xi}, the Hodge dual of $(df)$ is a Killing vector. Thence, $df$ is a closed-conformal Killing--Yano 3-form. As such, the derivative of this object is entirely captured by 
\be
\nu^{ab}=\nabla_c(df)^{abc}\,. 
\ee
In other words, the 
following 3 objects encode information about $f$, and its first, and second order derivatives:
\be
f_{ab}\,,\quad (df)_{abc}\,,\quad \nu_{ab}\,.
\ee
They provide alternatives to \eqref{k2der}, capturing the information about the Hodge dual of $k$ and its derivatives; we shall return to this possibility in App~\ref{App:f}.

We shall now discuss symmetry operators for various test field equations in a Kerr--NUT--AdS background spacetime. We seek these to be homogeneous in a number of derivatives and number of PKY tensors. In particular, since we consider wave-type equations that are second order in derivatives, we shall focus on homogeneous symmetry operators that contain exactly two covariant derivatives for each term, some possibly acting on the principal tensor. In the next section, we start by reviewing the well known scalar case. We then proceed to more complicated vector and tensor perturbations.

\section{Scalar field perturbations}\label{Sec:scalar}

\subsection{Separability of scalar field equation}

In a generic spacetime, there are only the following two objects available for a construction of the scalar field operator:\footnote{In principle, one could also add curvature terms. While such terms are important, for example for the conformal wave equation (see e.g. \cite{Gray:2020rtr,Gray:2021wzf}), in this paper we do not consider this possibility, neither for the field operators, nor for their symmetry operators. Whether the latter is a serious restriction remains to be seen in future.}  
\be\label{generic}
g^{ab}\,,\quad \nabla_a\,. 
\ee 
The only possibility for a 2nd-order in derivatives operator is thus
\be 
\nabla^2=g^{ab}\nabla_a \nabla_b\,. 
\ee
The (massive) scalar equation can then be written as an eigenvalue equation:
\be
\nabla^2\Phi=\mu^2\Phi\,, 
\ee
where $\mu$ stands for the mass of the scalar field.

{
It turns out that the above scalar field equation admits separability in the general off-shell canonical spacetimes \cite{Carter:1968cmp}. Namely, a solution can be found in the following separated form (e.g. \cite{Frolov:2017kze}):
\be
\Phi=R(r)Y(y) e^{-i\omega\tau+im\psi}\,,
\ee
where functions $R(r)$ and $Y(y)$ obey the following 2nd-order ODE's:
\ba
\partial_r(\Delta_r\partial_r R)+\frac{{\cal  X}_r}{\Delta_r}R=0\,,\nonumber\\
\partial_y(\Delta_y\partial_y Y)+\frac{{\cal X}_y}{\Delta_y}Y=0\,, 
\ea
where
\ba
{\cal X}_r&=& (\omega r^2-m)^2-\Delta_r(\kappa+\mu^2r^2)\,,\nonumber\\
{\cal X}_y&=& -(\omega y^2-m)^2+\Delta_y(\kappa-\mu^2y^2)\,,
\ea
and $\kappa$ is the separation constant. 
}

{Importantly, the above separability is possible due to the existence of a complete set of four mutually commuting operators, one of which is scalar wave operator $\nabla^2$, whose common eigenfunction is the separated solution. Namely, one has the following set: 
\be\label{setscalar}
\{\nabla^2, O^{(s)}_l, O^{(s)}_q, O^{(s)}_c\}\,, 
\ee
where 
\ba
 O^{(s)}_l&=&\xi^a\nabla_a\,,\nonumber\\
 O^{(s)}_q&=&\nabla_a(K^{ab}\nabla_b)\,,\nonumber\\
 O^{(s)}_c&=&\zeta^a\nabla_a\,,
\ea
are homogeneous symmetry operators that  are linear, quadratic, and cubic, respectively in the PKY tensor and contain exactly two derivatives.\footnote{In higher dimensions, the situation is even more remarkable, as in $d$ dimensions, one finds $(d-1)$ homogenous symmetry operators that contain $1, 2,\dots (d-1)$ powers of the PKY tensor $k$, see \cite{Frolov:2017kze} for more details.}
The separated solution above is then the common eigenfunction  of these operators: 
\ba
\nabla^2\Phi&=&\mu^2\Phi\,, \nonumber\\
 O^{(s)}_l\Phi&=&-i\omega \Phi\,,\nonumber\\
  O^{(s)}_q\Phi&=&\kappa \Phi\,,\nonumber\\
   O^{(s)}_c\Phi&=&im \Phi\,.
\ea
Obviously, the separation constant $\kappa$ is simply the eigenvalue of the quadratic in $k$ operator  $O^{(s)}_q$.
}

\subsection{Homogeneous symmetry operators for scalar fields} 
 
{ 
Let us now show how the above symmetry operators can be systematically derived by seeking the homogeneous operators that contain two derivatives and are linear, quadratic, and cubic in the PKY tensor.  
To construct such operators, we can use the following objects:
\be
g^{ab}\,,\ \nabla_a\,,\ 
k^{ab}\,,\ \xi^a\,,\ (d\xi)_{ab}\,, 
\ee  
where the last three are uniquely constructed from $k$ by applying zero, one, and two derivatives. To simplify our notation let us introduce the following shorthands:
\ba
\nabla_\xi&=&\xi^a\nabla_a\,,\quad \nabla_\zeta=\zeta^a \nabla_a\,,\nonumber\\
\nabla^2_k&=&k^{ab}\nabla_a\nabla_b\,,\quad \nabla^2_{k^2}=(k^2)^{ab}\nabla_a\nabla_b\,,\nonumber\\
\nabla^2_{k^3}&=&(k^3)^{ab}
\nabla_a\nabla_b\,.
\ea 
Of course, since there is no torsion, $\nabla^2_k$ and $\nabla^2_{k^3}$ both vanish on scalars, but may be non-trivial when applied to vectors and tensors. 
}

\subsubsection{Linear in $k$ operators}

We start with linear in $k$ scalar operators, starting with no derivatives on $k$, then one derivative, and finally two. As discussed above, since there is no torsion, we have to discard $\nabla^2_k$, which is the only possibility including $k$. 
The only operator containing 1 derivative on $k$ is
\be
O^{(s)}_l=\nabla_\xi={\cal L}_{\xi}=\partial_\tau\,. 
\ee 
Moreover, there is no scalar operator with two derivatives on $k$, so $O^{(s)}_l$ is the only linear operator at our disposal. One can easily show that it commutes with the box operator:
\be
[O^{(s)}_l, \nabla^2]=0\,. 
\ee 

\subsubsection{Quadratic in $k$ operators}
Lets now proceed to operators with two $k$'s. Then we have the following possibilities:
\ba
O_q^1&=&\nabla^2_{k^2}\,,\quad O_q^2=k^2\nabla^2\,,\\
O_q^3&=&k^a{}_b\xi^b \nabla_a\,,\\
O_q^4&=&\xi^2\,,\quad O_q^5=k_{ab}(d\xi)^{ab}\,.
\ea 
The potential most general homogeneous operator quadratic in $k$ is thus 
\be
O_q=
\alpha_1 O_q^1+\alpha_2 O_q^2+
\alpha_3 O_q^3+\alpha_4 O_q^4+\alpha_5 O_q^5\,.
\ee
We now require that 
\be
[O_q,\nabla^2]=0\,, 
\ee
for any off-shell functions $\Delta_r$ and $\Delta_y$. 
This implies the following constraints:
\be
\alpha_2=\frac{\alpha_1}{2}\,,\quad \alpha_3=-2\alpha_1\,,\quad \alpha_4=0\,,\quad \alpha_5=0\,.
\ee
So, we have a 1-parametric family of quadratic in $k$ operators. Setting $\alpha_1=1$ w.l.o.g., the most general commuting operator thus reads 
\be
O_q^{(s)}= \nabla^2_{k^2}+\frac{1}{2}k^2\nabla^2-2k^a{}_b\xi^b \nabla_a\,.
\ee
In fact, this is nothing else than the well known operator, constructed a long time ago by Carter \cite{Carter:1977pq}, as we have 
\begin{align}
O_q^{(s)}&=
\nabla_a( k^a{}_ck^{cb}+\frac{1}{2} k^2g^{a b}\nabla_b )
\nonumber
\\ 
&-\nabla_a(k^a{}_ck^{cb}+\frac12 k^2 g^{a b})\nabla_b -2k^a{}_b\xi^b \nabla_a
\nonumber
\\
&=\nabla_a(K^{a b}\nabla_b )\,,
\end{align}
where in the final step we have made use of \eqref{PKYT} and \eqref{CKT and KT explicit expressions} to recover the usual second order symmetry operator corresponding to Killing tensor $K^{ab}$.

\subsubsection{Cubic in $k$ operators}
Let us finally proceed to cubic in $k$ operators. Since on scalars $\nabla^2_{k^3}$ vanishes, and we find 
\be
(k^2)^{ab}(d\xi)_{ab}=0\,,
\ee
we have the following possibilities:
\ba
O_c^1&=&(k^2)_{ac}\xi^c \nabla^a\,,\quad 
O_c^2=k^2 \xi^a\nabla_a\,, 
\ea
{which leads to the most general linear combination 
\be
O_c=\alpha_1 O_c^1+\alpha_2 O_c^2\,.
\ee
}
The requirement that this commutes with $\nabla^2$ then yields
\be
\alpha_2=\frac{\alpha_1}{2}\,. 
\ee  

Thus, the most general symmetry operator of this order is given by 
\be
O_c^{(s)}= (k^2)_{ac} \xi^c \nabla^a+\frac{1}{2}k^2 \xi^a\nabla_a\,.
\ee 
Of course, this operator is also well known, as we have 
\be 
O_c^{(s)}=K^{a}{}_b\xi^b\nabla_a= \zeta^a\nabla_a={\cal L}_{\zeta}=
\partial_\phi\,,
\ee
on behalf of \eqref{zeta}.

It is easy to show that all the above symmetry operators mutually commute. We have thus systematically recovered the complete set \eqref{setscalar} of mutually commuting operators that intrinsically characterize the separability of scalar perturbations in the off-shell Kerr--NUT--AdS spacetimes.\footnote{
This procedure also applies to the construction of the symmetry operators of the conformal wave equation~\cite{Gray:2021wzf}, where the additional terms present are also homogeneous in $k$. For brevity, we do not repeat the corresponding construction here, and refer the interested reader to \cite{Gray:2021wzf}.}

\section{Vector perturbations}\label{Sec:vector}
\subsection{Vector operator} 
{In this section we shall seek homogeneous symmetry operators for (massive) vector perturbations. The corresponding field strength $F_{ab}$ is determined from a vector potential $A^a$ by
\be
F=dA\,, 
\ee
i.e. $F_{ab}=\nabla_a A_b-\nabla_b A_a$.
In what follows we directly work with the vector potential $A^a$. Since such a field carries a vector index, we seek `matrix' operators $O^a{}_b$ that act on $A^a$ as follows:
\be
(OA)^a\equiv O^a{}_b A^b\,.   
\ee 
} 

{In a generic spacetime, where the only available objects are\footnote{{In principle, one can also consider a Hodge star operation which, however, changes rank of the operator as the number of dimensions are varied. For example, in 4d one could consider an operator
\be
\epsilon^{a}{}_b{}^{cd}\nabla_c\nabla_d\,, 
\ee
where $\epsilon^{abcd}$ stands for the Levi-Civita tensor. 
In what follows we do not consider this possibility, see however App.~\ref{App:f}.}
}   
\be
\nabla_a\,,\quad g^{ab}\,,\quad \delta^a_b\,,\quad g_{ab}\,, 
\ee
we have the following three possibilities for operator $O$:
\be
\nabla^2 \delta^a_b\,,\quad \nabla_b\nabla^a\,,\quad \nabla^a\nabla_b\,.
\ee
(Note that the latter two coincide in vacuum spacetimes.) The {\em vector operator} uses the first two and reads 
\be\label{VO}
(O_M)^a{}_b =\nabla^2 \delta^a_b-\nabla_b \nabla^a\,,
\ee   
acting on the vector potential as
\be
(O_M)^a{}_b A^b=\nabla_b F^{ba}=\nabla^2 A^a-  \nabla_b \nabla^a A^b\,. 
\ee
}

{
The massive vector equation can then be written as an eigenvalue equation:
\be
(O_M)^a{}_b A^b=\mu^2 A^a\,. 
\ee
Taking its divergence, one recovers the corresponding consistency condition:
\be\label{Lorentz}
\nabla_a A^a=0\,. 
\ee
When $\mu=0$, which corresponds to the Maxwell case, the theory is a gauge theory, and the latter equation is not a consequence of the equations of motion, though it can be imposed as a Lorenz gauge fixing condition.  
Herein we shall not explicitly impose \eqref{Lorentz}.
}

{To find the symmetry operators for  the vector operator \eqref{VO}, that is to say operators $O$ satisfying 
\be\label{Vcomm}
[O,O_M]^a{}_{b}=O^a{}_c (O_M)^c{}_b-(O_M)^a{}_c O^c{}_b=0\,, 
\ee
we proceed in a way similar to the scalar case, except we have now more possibilities due to the fact that the field carries a non-trivial index. 
We shall also distinguish two cases -- the off-shell canonical case and the vacuum (possibly with $\Lambda$) Kerr--NUT--AdS case.
We start with linear in $k$ operators. 
}

\subsection{Linear in $k$ operators}
We have the following options for linear in $k$ matrix [i.e. (1,1)-tensor] operators {(for notation simplicity not writing the matrix indices on operators)}:
\ba
O_1&=&k^a{}_b \nabla^2\,,\quad 
O_2= k_{cb}\nabla^a \nabla^c\,,\quad 
O_3= k_{cb}\nabla^c \nabla^a\,,
\nonumber\\
O_4&=& \nabla^2_k \delta^a_b\,,\quad
O_5=k^{ac}\nabla_c\nabla_b\,,\quad 
O_6=k^{ac}\nabla_b\nabla_c
 \nonumber\\
O_7&=&
 \nabla_\xi \delta^a_b\,,\quad 
O_8=\xi_b \nabla^a\,,\quad 
O_9=\xi^a \nabla_b\,,\nonumber\\ 
O_{10}&=&(d\xi)^a{}_b\,.
\ea

Let us first focus on the off-shell canonical spacetimes.\footnote{In the off-shell case, we assume that the metric functions $\Delta_r(r)$ and $\Delta_y(y)$ are sufficiently general, that is, several of their lower derivatives are non-zero and can be treated as `independent'. This will be true also on-shell for dimensions higher than six, but does not continue to hold on-shell in four dimension. Hence the need to distinguish the cases.}

{In what follows, as before in the scalar case, for both vector and tensor perturbations we will always use $\alpha_i$ to denote the constant coefficients of the most general linear combination of possible operators, of a given degree. That is, schematically (suppressing the respective indices) we consider the general operators
\be
O_{\text{Gen}}=\sum_{i}\alpha_i O_i\,.
\ee
}
By requiring commutativity with the Maxwell operator, we find the following 6 constraints:
\ba
\alpha_3&=&-\alpha_1-\alpha_2\,,\quad \alpha_5=-\alpha_2+\alpha_4-\alpha_1\,,\nonumber\\
\alpha_6&=&\alpha_2-\alpha_4\,,\quad \alpha_8=-2\alpha_1\,,\quad \alpha_9=0\,,\nonumber\\
\alpha_{10}&=&\alpha_2-\alpha_4+\frac{1}{2}\alpha_7\,.
\ea
This means that in principle we have 4 operators of this kind that commute with the Maxwell operator. However, {there is a 2-parametric family of trivial operators, characterized by an arbitrary choice of $\{\alpha_2, \alpha_4\}$. In what follows we eliminate this freedom by setting
\be
\alpha_2=\alpha_4=0\,. 
\ee
}
We are just left with 2 independent non-trivial symmetry operators, characterized by $\{\alpha_1, \alpha_7\}$. Let us write their representatives. First, setting $\alpha_1=1$ and $\alpha_7=0$, we have 
\be\label{newlinear}
O_k^{(v)}= k^a{}_b \nabla^2-k_{cb}\nabla^c \nabla^a
-k^{ac}\nabla_c\nabla_b-2\xi_b \nabla^a\,.
\ee
{This operator remains non-trivial upon imposing \eqref{Lorentz}.}
Second, setting $\alpha_1=0$ and $\alpha_7=1$ and the remaining ones to zero, we recover 
\ba
O^{(v)}_\xi&=&\xi^c \nabla_c \delta^a_b+\frac{1}{2}(d\xi)^a{}_b={\cal L}_{\xi}\delta^a_b=\delta^a_b \partial_t\,.
\ea 
This is thus an operator corresponding to the Lie derivative along the primary Killing vector $\xi$.

When the vacuum ($\Lambda=0$) on-shell Kerr--NUT spacetime is considered, we find that the requirement for commutativity results only in 5 constraints:
\ba
\alpha_3&=&-\alpha_1-\alpha_2\,,\quad 
\alpha_6=-\alpha_1-\alpha_5
\nonumber\\
\alpha_8&=&-2\alpha_1\,,\quad \alpha_9=0\,,\nonumber\\
\alpha_{10}&=&\alpha_2-\alpha_4+\frac{1}{2}\alpha_7\,.
\ea
However, in this case we find that there is a 3-parametric family of trivial operators,
characterized by arbitrary choice of $\{\alpha_2, \alpha_4, \alpha_5\}$. To {eliminate these, we could for example set
\be
\alpha_2=\alpha_4=0\,,\quad \alpha_5=-\alpha_1\,. 
\ee
So again we are left with 2 independent symmetry operators, characterized by $\{\alpha_1,\alpha_7\}$ -- proceeding as above, we then formally recover `the same' symmetry operators as in the off-shell case.
}

\subsection{Quadratic in $k$ operators}
Let us now turn to finding quadratic in PKY tensor homogeneous operators for the vector perturbations. We have the following 22 possibilities:
\ba
O_1&=&k^a{}_b \nabla_k^2\,,\quad 
O_2=k^a{}_c k_{bd} \nabla^c \nabla^d\,,\nonumber\\ 
O_3&=&k^a{}_c k_{bd}\nabla^d \nabla^c\,, \quad 
O_4=\nabla^2_{k^2}\delta^a_b\,,\nonumber\\
O_5&=&(k^2)^a{}_e \nabla^e \nabla_b\,,\quad 
 O_6=(k^2)^a{}_e \nabla_b \nabla^e\,,\nonumber\\
 O_7&=&(k^2)_{be} \nabla^e \nabla^a\,,\quad 
 O_8=(k^2)_{be} \nabla^a \nabla^e\,,\nonumber\\
O_9&=&k^2\nabla^2 \delta^a_b\,,\quad 
O_{10}=k^2 \nabla^a\nabla_b\,,\nonumber\\
O_{11}&=&k^2 \nabla_b \nabla^a\,,\quad 
O_{12}=(k^2)^a{}_b \nabla^2\,,\nonumber\\  
O_{13}&=&k^{a}{}_b \nabla_\xi\,,\quad 
O_{14}=k^e{}_f \xi^f \nabla_e 
\delta^a_b\,,\\ 
O_{15}&=&\xi^a k_{be}\nabla^e\,,\quad 
O_{16}=\xi_b k^a{}_e \nabla^e\,,\nonumber\\
O_{17}&=&k^a{}_c \xi^c \nabla_b\,,\quad 
O_{18}=k_{eb} \xi^e \nabla^a\,, \nonumber\\
O_{19}&=&(d\xi)^{ae}k_{be}\,,\quad 
O_{20}=(d\xi)_{be}k^{ae}\,,\nonumber\\
O_{21}&=&\xi^2 \delta^a_b\,,\quad 
O_{22}=\xi^a\xi_b\,. \nonumber
\ea
We note that not all of these are independent however for the purposes of counting it is useful to list everything in this manner. 

{
Focusing first on the off-shell Kerr--NUT--AdS spacetimes,} we find that the requirement for commutation \eqref{Vcomm} yields 16 constraints on the linear combination of the above operators: 
\ba 
\alpha_1&=&\alpha_3+2\alpha_{11}+\alpha_{16}+
\alpha_{19}+\alpha_{20}\,,\nonumber\\
\alpha_2&=&-\alpha_3-4\alpha_{11}-\alpha_{16}\,,\quad \alpha_4=-2\alpha_{11}\,,\nonumber\\
\alpha_5&=&2\alpha_{11}+\alpha_{16}+\alpha_{19}+\alpha_{20}\,,\nonumber\\
\alpha_6&=&2\alpha_{11}-\alpha_{19}-\alpha_{20}\,,\quad \alpha_{7}=2\alpha_{11}\,,\nonumber\\
\alpha_8&=&2\alpha_{11}+\alpha_{16}\,,\quad
\alpha_{9}=-\alpha_{11}\,,\quad 
\alpha_{12}=-4\alpha_{11}-\alpha_{16}\,,\nonumber\\
\alpha_{13}&=&-4\alpha_{11}-3\alpha_{16}\,,\quad \alpha_{14}=-\alpha_{16}\,,\quad 
\alpha_{15}=4\alpha_{11}\,,\nonumber\\
\alpha_{17}&=&4\alpha_{10}\,,\quad \alpha_{18}=8 \alpha_{11}+3\alpha_{16}\,,\nonumber\\
\alpha_{21}&=&-4\alpha_{11}-2\alpha_{16}\,,\quad 
\alpha_{22}=4\alpha_{11}+2\alpha_{16}\,.
\ea
Moreover, it can be shown that we have a 3-parametric {family of trivial operators, encoded in arbitrary $\{\alpha_3, \alpha_{19}, \alpha_{20}\}$. We can eliminate these by setting}
\be
\alpha_3=\alpha_{19}=\alpha_{20}=0\,. 
\ee
This means that we are left with 3 independent quadratic operators, characterized by specifying $\{\alpha_{10}, \alpha_{11}, \alpha_{16}\}$.  We write these as follows. 

First,  setting
\be
\alpha_{10}=1\,,\quad \alpha_{11}=0=\alpha_{16}\,, 
\ee 
we recover 
the following commuting operator:
\be
O^{(v)}_{q_1}=k^2\nabla^a\nabla_{b}+4k^a{}_c\xi^c \nabla_b=\nabla^a(k^2\nabla_b)\,.
\ee
This operator will vanish for the Proca field, or upon imposing the Lorenz gauge condition. 
In the vacuum Kerr case (with $\Lambda=0$ and no NUT parameter), this operator agrees with the operator 
${\cal K}_2$ in \cite{Mei:2023pho}.

Next, requiring that the absolute term is missing, we set
\be
\alpha_{10}=0\,,\quad \alpha_{11}=-\frac{1}{2}\,,\quad \alpha_{16}=1\,, 
\ee
to obtain the following operator:
\ba
O_{q_2}^{(v)}&=&
k^a{}_c k_{bd} \nabla^c \nabla^d
+\nabla^2_{k^2}\delta^a_b-(k^2)^a{}_e \nabla_b \nabla^e\nonumber\\
&&-(k^2)_{be} \nabla^e \nabla^a
+\frac{1}{2}k^2\nabla^2 \delta^a_b
-\frac{1}{2}k^2 \nabla_b \nabla^a\nonumber\\
&&+(k^2)^a{}_b \nabla^2-k^{a}{}_b \nabla_\xi
-k^e{}_f \xi^f \nabla_e 
\delta^a_b\nonumber\\
&&-2\xi^a k_{be}\nabla^e+\xi_b k^a{}_e \nabla^e-k_{eb} \xi^e \nabla^a\,.
\ea
This operator is gauge invariant and  in the vacuum case (again with no NUT parameter)
identical to the operator ${\cal K}_4$ found recently in \cite{Mei:2023pho}.

Finally, setting 
\be
\alpha_{10}=0=\alpha_{11}\,,\quad \alpha_{16}=1\,, 
\ee
we recover the following operator:
\ba
O_{q_3}^{(v)}&=&k^a{}_b \nabla_k^2-
k^a{}_c k_{bd} \nabla^c \nabla^d+
(k^2)^a{}_e \nabla^e \nabla_b\nonumber\\
&&+(k^2)_{be} \nabla^a \nabla^e
-(k^2)^a{}_b \nabla^2
-3k^{a}{}_b \nabla_\xi\nonumber\\
&&-k^e{}_f \xi^f \nabla_e 
\delta^a_b+\xi_b k^a{}_e \nabla^e
+3k_{eb} \xi^e \nabla^a\nonumber\\
&&-2\xi^2 \delta^a_b+2\xi^a\xi_b\,.
\ea
Both these operators, $O_{q_2}^{(v)}$ and $O_{q_3}^{(v)}$ will not vanish upon imposing $\nabla_aA^a=0$.

{Let us next turn to the on-shell vacuum ($\Lambda=0$) Kerr--NUT spacetimes.} In this case,  the operators $O_{10}$ and $O_{11}$ are no longer independent, and the above construction only yields two commuting operators. However, we can repeat the whole construction again, with the following results. The requirement for commutation yields 13 constraints: {
\ba
 \alpha_1&=&\alpha_3+2\alpha_8+\frac{1}{2}(\alpha_{12}-\alpha_{16})+\alpha_{19}+\alpha_{20}\,,\nonumber\\
 \alpha_{2}&=&-\alpha_3+\alpha_{12}\,,\quad \alpha_4=\frac{1}{2}(\alpha_{12}+\alpha_{16})\,,\nonumber\\
 \alpha_5&=&-\alpha_6-\alpha_{12}\,,\quad \alpha_7=-\alpha_8-\alpha_{12}\,,\nonumber\\
 \alpha_9&=&\frac{1}{4}(\alpha_{12}+\alpha_{16})\,,\quad 
 \alpha_{13}=\alpha_{12}-2\alpha_{16}\,,\nonumber\\
 \alpha_{14}&=&-\alpha_{16}\,,\quad 
 \alpha_{15}=-\alpha_{12}-\alpha_{16}\,,\nonumber\\
 \alpha_{17}&=&4\alpha_{10}+4\alpha_{11}+\alpha_{12}+\alpha_{16}\,,\quad 
 \alpha_{18}=\alpha_{16}-2\alpha_{12}\,,\nonumber\\
 \alpha_{21}&=&-\alpha_8+\frac{1}{2}(\alpha_{12}-\alpha_{16})\,,\quad \alpha_{22}=\alpha_{16}-\alpha_{12}\,.
\ea
Moreover, one can show that there is a remaining 5-parametric family of trivial (or identical) operators, encoded in 
$\{\alpha_3,\alpha_6, \alpha_8, \alpha_{19}, \alpha_{20}\}$.  To eliminate these, we may set
\be
\alpha_3=\alpha_6=\alpha_8=\alpha_{19}=\alpha_{20}=0\,. 
\ee
At the same time, since $\alpha_{10}$ and $\alpha_{11}$ are now not independent, we can also set 
\be
\alpha_{11}=0\,.
\ee
Thus, similar to \cite{Mei:2023pho}, we are left with a 3-parametric family of (non-trivial) commuting operators. In our case these are characterized by a choice of 
\be 
\{\alpha_{10}\,,\alpha_{12}\,,\alpha_{16}\}\,.
\ee
In particular, the operator ${\cal K}_2$ in 
\cite{Mei:2023pho} is reproduced upon setting 
$
\alpha_{10}=1\,\ \alpha_{12}=\alpha_{16}=0\,, 
$
and the operator ${\cal K}_4$ corresponds to 
$
\alpha_{10}=-\frac{1}{2}\,,\ \alpha_{12}=\alpha_{16}=1\,. 
$
We note these remain generalizations of the operators in~\cite{Mei:2023pho} since they contain non-trivial NUT parameter. Finally, setting 
\be
\alpha_{10}=\alpha_{12}=0\,,\quad \alpha_{16}=-2\,, 
\ee
we recover 
\ba
\tilde O_{q_3}^{(v)}&=& 
k^a{}_b \nabla_k^2-
\nabla^2_{k^2}\delta^a_b
-\frac{1}{2}k^2\nabla^2\delta^a_b\nonumber\\
&&+4k^a{}_b\nabla_\xi+2k^e{}_f\xi^f\nabla_e \delta^a_b+2\xi^ak_{be}\nabla^e\nonumber\\
&&-2\xi_b k^a{}_e \nabla^e-2k^a{}_c\xi^c \nabla_b-2k_{eb}\xi^e \nabla^a\nonumber\\
&&+\xi^2 \delta^a_b-2\xi^a\xi_b\,,
\ea
completing the set of the on-shell commuting operators. 
}

\subsection{Cubic in $k$ operators}
{We finally examine the cubic in $k$ vector operators of degree two. We organize these according to how many derivatives act on the (three factors of the) principal tensor. We start with operators with no derivatives, proceed to operators with one derivative, and finalize with two derivatives. In this way we find a total of 34 operators listed below. Note, however, 
that not all of these operators are necessarily independent.}

{
Let us start with operators where no derivatives act on the three copies of the PKY tensor, that is operators of the symbolic  ``$k^3\nabla^2$" form. These are given by:}
\ba
O_{1}&=&(k^2)^a{}_b \nabla^2_k\,,\quad 
O_{2}= k^{a}{}_b \nabla^2_{k^2}\,,\nonumber\\
O_{3}&=&\nabla^2_{k^3} \delta^a_b\,,\quad  
O_{4}=k^a{}_c (k^2)_{be}\nabla^c\nabla^e\,,\nonumber\\
O_{5}&=&k^a{}_c (k^2)_{be}\nabla^e \nabla^c\,,\quad 
O_{6}=(k^2)^a{}_c k_{be}\nabla^c \nabla^e\,,\nonumber\\
O_{7}&=&(k^2)^a{}_ck_{be}\nabla^e\nabla^c\,,\quad
O_{8}=k^2\nabla^2_k \delta^a_b\,,\nonumber\\
O_{9}&=&(k^3)^a{}_e\nabla^e\nabla_b\,,\quad O_{10}=(k^3)^a{}_e\nabla_b\nabla^e\,,\nonumber\\
O_{11}&=&(k^3)_{be}\nabla^e\nabla^a\,,\quad O_{12}=(k^3)_{be}\nabla^a\nabla^e\,,\\
O_{13}&=&k^2k^a{}_e\nabla^e\nabla_b\,,\quad O_{14}=k^2k^a{}_e\nabla_b\nabla^e\,,\nonumber\\
O_{15}&=&k^2 k_{be}\nabla^e\nabla^a\,,\quad 
O_{16}=k^2k_{be}\nabla^a\nabla^e\,,\nonumber\\
O_{17}&=&(k^3)^a{}_b \nabla^2\,,\quad O_{18}=k^a{}_b k^2 \nabla^2\,.\nonumber 
\ea
Note that the operators $\{O_1,\dots, O_8\}$ have two contractions between the $k$'s and the covariant derivative operators, the operators $\{O_9, \dots, O_{16}\}$ have one, and the operators $O_{17}, O_{18}$ have zero contractions. 

Next, we consider operators with one derivative acting on $k$'s, that is, the  terms of the symbolic ``$k^2 \xi \nabla$'' form. They read:
\ba
O_{19}&=&(k^2)^a{}_b \xi^c \nabla_c\,,\quad 
O_{20}=k^2 \xi^c\nabla_c \delta^a_b\,,\nonumber\\
O_{21}&=&(k^2)^{ca}\xi_b\nabla_c\,,\quad 
O_{22}=(k^2)_{cb}\xi^a \nabla^c\,,\nonumber\\
O_{23}&=&(k^2)^{cd}\xi_d\nabla_c \delta^{a}{}_{b}\,,\nonumber\\
O_{24}&=&k^2 \xi^a\nabla_b\,,\quad 
O_{25}=k^2 \xi_b \nabla^a\,,\nonumber\\
O_{26}&=&(k^2)^a{}_c\xi^c \nabla_b\,,\quad
O_{27}= (k^2)_{bc}\xi^c \nabla^a\,.
\ea
Here, $\{O_{19},\dots O_{23}\}$ have one contraction between the $k$'s and the covariant derivatives, and the operators $\{O_{24},\dots, O_{27}\}$ have no such contraction. 

{Finally, we consider terms with two derivatives acting on $k$'s, that is terms of the symbolic ``$\xi\xi k$" or ``$(d\xi)kk$" form. These are:}
\ba
O_{28}&=&k^a{}_c \xi^c \xi_b\,,\quad 
O_{29}=k_{bc}\xi^c \xi^a\,,\quad 
O_{30}=\xi^2 k^a{}_b\,,\nonumber\\
O_{31}&=&(d\xi)^a{}_c (k^2)^c{}_b\,,\quad 
O_{32}=(d\xi)_{bc}(k^2)^{ca}\,,\nonumber\\
O_{33}&=&k^a{}_e (d\xi)^{ef}k_{fb}\,,\quad 
O_{34}=k^2 d\xi^a{}_b\,.
\ea

{
For the off-shell Kerr--NUT--AdS spacetimes
one can} check, through tedious computer algebra calculations, that the only non trivial cubic operator that commutes with the vector operator is the Lie derivative along the secondary Killing vector $\zeta$, explicitly given by
\begin{multline}
O^{(v)}_\zeta =O_{23}+\frac{1}{2}O_{20}+\frac{1}{4}( O_{31}-O_{32} )
\\
+\frac 1 4 O_{34} +(O_{28} - O_{29})-O_{30}\,,
\end{multline}
or, upon using the definitions of $\zeta$ \eqref{zeta}, $K$ \eqref{CKT and KT explicit expressions}, and $k$ \eqref{PKYT}:
\ba\label{eq: zeta op}
O^{(v)}_\zeta&=&\zeta^c \nabla_c \delta^a_b+\frac{1}{2}(d\zeta)^a{}_b={\cal L}_{\zeta}\delta^a_b=\delta^a_b \partial_\psi\,.
\ea  
A similar procedure shows that there are no additional non-trivial symmetry operators when we are on-shell, {neither with nor without $\Lambda$.}

\subsection{Vector symmetry operators: summary}

To summarize, for the off-shell Kerr--NUT--AdS spacetimes we have obtained the following 6 symmetry operators: 
\be
\{O^{(v)}_\xi, O^{(v)}_k, O_{q_1}^{(v)}, O_{q_2}^{(v)}, O_{q_3}^{(v)}, O_{\zeta}^{(v)}\}\,.   
\ee
Together with the vector operator $O_M$, these form a set of 7 mutually commuting 
operators. For the on-shell vacuum case, one has an analogous set of 7 symmetry operators.

It remains to be seen whether by considering the corresponding eigenvalue problem, one could obtain the associated R-separated solution for vector potential perturbations, as attempted in \cite{Mei:2023pho}.

\section{Tensor perturbations}\label{Sec:tensor}

\subsection{Lichnerowicz operator}
Let us finally turn to tensor perturbations,
\be
g_{ab}\to g_{ab}+h_{ab}\,. 
\ee
Then, to the linear order in the perturbation theory, we find the Lichnerowicz operator:
 \ba
 (Lh)_{ab}&=&\delta R_{ab}\\
 &=&\frac{1}{2}\nabla^c(\nabla_a h_{bc}+\nabla_{b}h_{ac}-\nabla_c h_{ab})-\frac{1}{2}\nabla_a\nabla_b h\,,\nonumber
\ea
where $h=h_{ab}g^{ab}$. 
That means that we can write the Lichnerowicz operator as 
\be
(Lh)_{ab}=L_{ab}{}^{cd}h_{cd}\,, 
\ee 
where 
\be
 L_{ab}{}^{cd}=
 \nabla^{(d} \nabla_{(a}\delta^{c)}{}_{b)}-\frac{1}{2}\nabla^2\delta^c_a\delta^d_b
 -\frac{1}{2}\nabla_{(a}\nabla_{b)} g^{cd}\,,
\ee
highlighting manifestly that $L$ is symmetric in both $(ab)$ and $(cd)$ indices. 

{In what follows, we seek symmetry operators for $L$, of the form
\be
(Oh)_{ab}\equiv O_{ab}{}^{cd} h_{cd}\, 
\ee
that are manifestly symmetric in $(ab)$ and $(cd)$ indices, and commute with $L$:
\be\label{Tcomm}
[O,L]_{ab}{}^{cd}=O_{ab}{}^{ef}L_{ef}{}^{cd}-L_{ab}{}^{ef}O_{ef}{}^{cd}=0\,.
\ee
To simplify our notation for the $O$'s, we assume but do not explicitly write  symmetrization  in $(ab)$ and $(cd)$ indices, and only restate it for the final result. 
}

\subsection{Linear operators}
{Let us first construct linear in $k$ tensor symmetry operators. We have the following 20 possibilities:}
\ba
O_1&=&\nabla^2_k \delta_a^c\delta_b^d\,,\quad 
O_2=\nabla_k^2 g^{cd} g_{ab}\,,\nonumber\\
O_3&=&k^{ec}\nabla_e \nabla^d g_{ab}\,,\quad 
O_4=k^{ec}\nabla^d \nabla_e g_{ab}\,,\nonumber\\
O_5&=&k^{ec}\nabla_e \nabla_a \delta^d_b\,,\quad 
O_6=k^{ec}\nabla_a \nabla_e \delta^d_b\,,\nonumber\\
O_7&=&k^e{}_a\nabla_e \nabla^c \delta^d_b\,,\quad 
O_8=k^e{}_a \nabla^c \nabla_e \delta^d_b\,,\nonumber\\
\label{eq: l ops Met 1 con 3}
O_9&=&k^e{}_a\nabla_e \nabla_b g^{cd}\,,\quad 
O_{10}=k^e{}_a \nabla_b \nabla_e g^{cd}\,, 
\label{eq: l ops Met 1 con 4}\nonumber\\
\label{eq: l ops Met 0 con 1}
O_{11}&=&k^c{}_a \nabla^2 \delta^d_b\,,\quad 
O_{12}=k^c{}_a \nabla^d\nabla_b\,,\nonumber\\ 
O_{13}&=&k^c{}_a\nabla_b\nabla^d\,,\quad
O_{14}=\nabla_\xi g^{cd}g_{ab}\,,\\
O_{15}&=&\nabla_\xi \delta^c_a \delta^d_b\,,\quad O_{16}=\xi^c \nabla^d g_{ab}\,,\nonumber\\
O_{17}&=&\xi^c\nabla_a \delta^d_b\,,\quad
O_{18}=\xi_a\nabla^c \delta^d_b\,,\nonumber\\
O_{19}&=&\xi_a \nabla_b g^{cd}\,,\quad 
O_{20}=(d\xi)^c{}_a \delta^d_b\,. \nonumber  
\ea
As always, not all of these operators are independent; in fact, the operator $O_2$  is trivial since it is the anti-symmetrized second derivative of a scalar.

{Let us first focus on the off-shell Kerr--NUT--AdS spacetimes. Then, 
by requiring the linear combination of these operators commutes with the Lichnerowicz operator, we find 15 constraints. However, 4 of the remaining linear combinations are trivial. We are thus left with only one commuting operator, the primary Killing vector operator:
\be\label{Oxit}
O^{(t)}_\xi=\nabla_\xi \delta^c_a \delta^d_b-
(d\xi)^{(c}{}_{(a} \delta^{d)}{}_{b)}=\delta^c_a \delta^d_b {\cal L}_{\xi}=\delta^c_a \delta^d_b\partial_t\,.
\ee
As we shall see, the situation is much more interesting in the on-shell case.
}

{
Namely, let us consider on-shell Kerr-NUT-AdS spacetimes with non-trivial $\Lambda$. Then,
by requiring that the linear combination of the above operators commutes with the Lichnerowicz operator, we find the following 12 constraints:
\ba
\alpha_4&=&-\alpha_3\,,\quad \alpha_6=-\alpha_1-\alpha_5\,,\quad \alpha_8=\alpha_1-\alpha_7\,,\nonumber\\ 
\alpha_9&=&-\alpha_1-\alpha_{10}-\alpha_{14}\,,\quad \alpha_{11}=-\alpha_1\,,\nonumber\\
\alpha_{12}&=&\alpha_1+\alpha_5+\alpha_7\,,\quad 
\alpha_{13}=\alpha_1-\alpha_5-\alpha_7+2\alpha_{14}\,,\nonumber\\
\alpha_{16}&=&-\alpha_1-2\alpha_{14}\,,\quad \alpha_{17}=-3\alpha_1\,,\quad 
\alpha_{18}=\alpha_1+2\alpha_{14}\,,\nonumber\\
\alpha_{19}&=&-\alpha_{14}\,,\quad 
\alpha_{20}=4\alpha_1+\alpha_5-\alpha_{15}\,.
\ea
Moreover, there is a 5-parametric family of trivial operators, characterized by $\{\alpha_2, \alpha_3, \alpha_5, \alpha_7, \alpha_{10}\}$; we eliminate these by setting 
\be
 \alpha_2=\alpha_3=\alpha_5=\alpha_7=\alpha_{10}=0\,.
\ee
We are just left with 3 non-trivial symmetry operators, characterized by $\{\alpha_1, \alpha_{14}, \alpha_{15}\}$. In particular, by setting 
\be
\alpha_{15}=1\,,\quad \alpha_1=\alpha_{14}=0\,, 
\ee
we recover the above Killing vector operator $O_\xi^{(t)}$, \eqref{Oxit}. Next, setting 
\be
\alpha_1=1\,,\quad \alpha_{14}=\alpha_{15}=0\,, 
\ee
we recover the following operator:
\ba 
O^{(t)}_{l_1}&=&\nabla^2_k \delta_a^c\delta_b^d-k^{e(c}\nabla_{(a} \nabla_{|e|} \delta^{d)}_{b)}+
k^e{}_{(a} \nabla^{(c} \nabla_{|e|} \delta^{d)}_{b)}\nonumber\\
&&-k^e{}_{(a}\nabla_{|e|} \nabla_{b)} g^{cd}
-k^{(c}{}_{(a} \nabla^2 \delta^{d)}_{b)}
+k^{(c}{}_{(a} \nabla^{d)}\nabla_{b)}\nonumber\\
&&+k^{(c}{}_{(a}\nabla_{b)}\nabla^{d)}-\xi^{(c }\nabla^{d)} g_{ab}-3\xi^{(c}\nabla_{(a} \delta^{d)}{}_{b)}\nonumber\\
&&+\xi_{(a}\nabla^{(c} \delta^{d)}{}_{b)}+4(d\xi)^{(c}{}_{(a} \delta^{d)}{}_{b)}\,.
\ea 
Finally, setting 
\be
\alpha_{14}=1\,, \quad \alpha_1=\alpha_{15}=0\,,
\ee
we have the following symmetry operator:
\ba
O^{(t)}_{l_2}&=&-k^e{}_{(a}\nabla_{|e|} \nabla_{b)} g^{cd}+2k^{(c}{}_{(a}\nabla_{b)}\nabla^{d)}+\nabla_\xi g^{cd}g_{ab}\nonumber\\
&&-2\xi^{(c} \nabla^{d)} g_{ab}+2\xi_{(a}\nabla^{(c} \delta^{d)}{}_{b)}-\xi_{(a} \nabla_{b)} g^{cd}\,.
\ea
Of course, both these symmetry operators remain symmetry operators also for the case of $\Lambda=0$, while no new independent operators are obtained in that case.
}

\subsection{Quadratic operators}

Let us next proceed to quadratic tensor operators. Denoting  
\be 
(d\xi\cdot k)=(d\xi)^{ab}k_{ab}\,,
\ee
we find {52 possible operators which we list below.} 

First, we have the operators of the type 
``$k^2\nabla^2$". These are given by: 
\ba
O_1&=&\nabla^2_{k^2} g_{ab}g^{cd}\,,\quad 
O_2=\nabla^2_{k^2}\delta^c_a \delta^d_b\,,\quad
O_3=k^c{}_a \nabla^2_{k} \delta^d_b\,,\nonumber\\
O_4&=& k_{ae}k_{bf}\nabla^e\nabla^f g^{cd}\,,\quad 
O_5=k^{ce}k^{df}\nabla_e\nabla_f g_{ab}\,,\nonumber\\
O_6&=&k^{ce}k_{af}\nabla_e \nabla^f \delta^d_b\,,\quad  
O_7=k^{ce}k_{af}\nabla^f \nabla_e \delta^d_b\,,\nonumber\\
O_8&=&(k^2)^{ce}\nabla_e \nabla^d g_{ab}\,,\quad  
O_9=(k^2)^{ce} \nabla^d \nabla_e g_{ab}\,,\nonumber\\
\label{eq:q ops Met  1 con 2}
O_{10}&=&(k^2)^{ce}\nabla_e\nabla_a\delta^d_b\,,\quad 
O_{11}=(k^2)^{ce}\nabla_a\nabla_e\delta^d_b\,,\nonumber\\
\label{eq:q ops Met  1 con 3}
O_{12}&=&(k^2)_{ae}\nabla^e\nabla_bg^{cd}\,,\quad 
O_{13}=(k^2)_{ae}\nabla_b\nabla^eg^{cd}\,,\nonumber\\
\label{eq:q ops Met  1 con 4}
O_{14}&=&(k^2)_{ae}\nabla^e\nabla^c\delta^d_b\,,\quad 
O_{15}=(k^2)_{ae}\nabla^c\nabla^e\delta^d_b\,,\nonumber\\
\label{eq:q ops Met 1 con 5}
O_{16}&=&k^c{}_a k^d{}_e \nabla^e \nabla_b\,,\quad 
O_{17}=k^c{}_a k^d{}_e \nabla_b \nabla^e\,,\nonumber\\
O_{18}&=&k^2 \nabla^2 g^{cd}g_{ab}\,,\quad O_{19}=k^2\nabla^2\delta^c_a \delta^d_b\,,
\label{eq: q ops Met 0 con 1}
\nonumber\\
O_{20}&=&k^c{}_a k^d{}_b \nabla^2\,,\quad 
O_{21}=(k^2)^c{}_a \nabla^d \nabla_b\,,
\label{eq: q ops Met 0 con 2}
\\
O_{22}&=& (k^2)^c{}_a \nabla_b \nabla^d\,,\quad 
O_{23}=
(k^2)_a{}^{c}\nabla^2 \delta_{b}{}^d\,,\nonumber
\label{eq: q ops Met 0 con 3}\\
O_{24}&=&k^2\nabla_a\nabla^c \delta_b^d\,,\quad 
O_{25}=k^2\nabla^c\nabla_a\delta_b^d\,,\nonumber\\
O_{26}&=&k^2\nabla_a\nabla_b g^{cd}\,,\quad O_{27}=k^2g_{ab}\nabla^c\nabla^d\,.\nonumber
\ea
Here, the operators $\{O_1,\dots O_7\}$ have two contractions between $k$'s and $\nabla$'s, the operators $\{O_8,\dots O_{17}\}$ have one, and operators $\{O_{18}, \dots O_{27}\}$ have no such contractions.

{
We then move to operators of the type ``$k \xi \nabla$"
\ba
O_{28}&=&
k^c{}_a \nabla_\xi \delta^d_b\,,\quad 
O_{29}=\xi^c k^d{}_e \nabla^e g_{ab}\,,\nonumber\\
O_{30}&=&\xi^c k_{ae} \nabla^e \delta^d_b\,,\quad
O_{31}=\xi_a k^c{}_e \nabla^e \delta^d_b\,,\nonumber\\
O_{32}&=&\xi_a k_{be}\nabla^e g^{cd}\,,\quad 
O_{33}=k^f{}_e \xi^e \nabla_f g^{cd}g_{ab}\,,\nonumber\\ 
O_{34}&=&k^f{}_e \xi^e \nabla_f \delta^c_a\delta^d_b\,, \nonumber\\
O_{35}&=&\xi^c k^d{}_a \nabla_b\,,\quad 
O_{36}=\xi_a k^c{}_b \nabla^d\,,\nonumber\\
O_{37}&=&k^c{}_e \xi^e \nabla^d g_{ab}\,,\quad 
O_{38}=k^c{}_e \xi^e \nabla_a \delta^d_b\,,\nonumber\\
O_{39}&=&k_{ae}\xi^e \nabla_b g^{cd}\,,\quad O_{40}=k_{ae}\xi^e \nabla^c \delta^d_b\,. 
\ea
Here, the operators $\{O_{28}, \dots, O_{34}\}$ have 1 contraction between $k$'s and $\nabla$, and operators $\{O_{35},\dots, O_{40}\}$ have no such contractions.
}

{
Finally, we have operators of the type ``$\xi\xi$" and ``$k (d\xi)$". These read:}
\ba
O_{41}&=&(d\xi\cdot k) g^{cd}g_{ab}\,,\quad  
O_{42}=(d\xi\cdot k) \delta^c_a \delta^d_b\,,\nonumber\\
O_{43}&=&(d\xi)^c{}_e k^{ed}g_{ab}\,,\quad 
O_{44}=(d\xi)^c{}_e k^e{}_a \delta^d_b\,,\nonumber\\
O_{45}&=&(d\xi)_{ae}k^{e}{}_b g^{cd}\,,\quad O_{46}=(d\xi)_{ae}k^{ec} \delta^d_b\,,\nonumber\\
O_{47}&=&(d\xi)^c{}_a k^{d}{}_b\,,\quad
O_{48}=\xi^c \xi^d g_{ab}\,,\nonumber\\ 
O_{49}&=&\xi^c \xi_a \delta^d_b\,,\quad 
O_{50}=\xi_a\xi_b g^{cd}\,,\nonumber\\
O_{51}&=&\xi^2 g^{cd}g_{ab}\,,\quad 
O_{52}=\xi^2 \delta^c_a \delta^d_b\,.
\ea
{Note that we have $(d\xi)_{a}{}^c k_{c b}=(d\xi)_{b}{}^{c} k_{c a}$, so such operators will not be trivial on symmetrization over the relevant indices.}

{Interestingly, we find that there are no  non-trivial symmetry operators for the off-shell Kerr-NUT-AdS spacetimes of this type.}

{
Moving on to the on-shell spacetimes with non-trivial $\Lambda$, we find 42 constraints and 7-parametric family of trivial operators. We are thus left with 3 non-trivial symmetry operators, in our case characterized by a choice of $\{\alpha_2, \alpha_{26}, \alpha_{37}\}$. First, setting 
\be
\alpha_{26}=1\,,\quad \alpha_2=0=\alpha_{37}\,, 
\ee
we find the following operator:
\ba
O^{(t)}_{q_1}&=&-2k^2\nabla_{(a}\nabla^{(c} \delta^{d)}{}_{b)}+
k^2\nabla_{(a}\nabla_{b)} g^{cd}
\nonumber\\
&&+4k_{(a|e|}\xi^e \nabla_{b)} g^{cd}
-8k_{(a|e|}\xi^e \nabla^{(c} \delta^{d)}{}_{b)}\,.
\ea
Next, setting 
\be
\alpha_{37}=2\,,\quad \alpha_2=\alpha_{26}=0\,, 
\ee
we find
\ba
O^{(t)}_{q_2}&=& 
(k^2)_{(a|e|}\nabla_{b)}\nabla^eg^{cd}
-2(k^2)^{(c}{}_{(a} \nabla_{b)} \nabla^{d)}\nonumber\\
&&-\xi_{(a} k_{b)e}\nabla^e g^{cd}
-k^f{}_e \xi^e \nabla_f g^{cd}g_{ab}\nonumber\\
&&-2\xi_{(a} k^{(c}{}_{b)} \nabla^{d)}
+2k^{(c}{}_e \xi^{|e|} \nabla^{d)} g_{ab}\nonumber\\
&&-k_{(a|e|}\xi^e \nabla_{b)} g^{cd}
+2k_{(a|e|}\xi^e \nabla^{(c} \delta^{d)}{}_{b)}\,.
\ea
Finally, setting 
\be
\alpha_2=1\,,\quad \alpha_{26}=\alpha_{37}=0\,, 
\ee
we have 
\ba
O^{(t)}_{q_3}&=& 
\nabla^2_{k^2}\delta^c_a \delta^d_b
-2(k^2)^{(c|e|}\nabla_e\nabla_{(a}\delta^{d)}{}_{b)}\nonumber\\
&&+2(k^2)^{(c|e|}\nabla_{(a}\nabla_{|e|}\delta^{d)}{}_{b)}
+\frac{1}{2}k^2\nabla^2\delta^c_a \delta^d_b\nonumber\\
&&+k^2\nabla_{(a}\nabla^{(c} \delta^{d)}{}_{b)}
-k^2\nabla^{(c}\nabla_{(a}\delta^{d)}{}_{b)}\nonumber\\
&&+8k^{(c}{}_{(a} \nabla_\xi \delta^{d)}{}_{b)}
+4\xi^{(c} k_{(a|e|} \nabla^e \delta^{d)}_{b)}\nonumber\\
&&-4\xi_{(a} k^{(c}{}_{|e|} \nabla^{|e|} \delta^{d)}{}_{b)}
-2k^f{}_e \xi^e \nabla_f \delta^c_a\delta^d_b\nonumber\\
&&-4k^{(c}{}_e \xi^{|e|} \nabla_{(a} \delta^{d)}{}_{b)}
+4k_{(a|e|}\xi^e \nabla^{(c} \delta^{d)}{}_{b)}\nonumber\\
&&-(d\xi)_{(a|e|}k^{e}{}_{b)} g^{cd}
+3(d\xi)_{(a|e|}k^{e(c} \delta^{d)}{}_{b)}\nonumber\\
&&-(d\xi)^{(c}{}_{(a} k^{d)}{}_{b)}
-2\xi^{(c} \xi^{d)} g_{ab}
+8\xi^{(c} \xi_{(a} \delta^{d)}{}_{b)}\nonumber\\
&&
-2\xi_a\xi_b g^{cd}+2\xi^2 g^{cd}g_{ab}
-6\xi^2 \delta^c_a \delta^d_b\,.
\ea
}

{Of course, all three of these operators remain also valid in the vacuum, $\Lambda=0$, case. Then, and further in the absence of the NUT charge, the operators $O^{(t)}_{q_1}$ and $O^{(t)}_{q_2}$ coincide with ${\cal K}_{3}$ and ${\cal K}_{2}$ respectively in \cite{Mei:2023pho}, while the operator $O^{(t)}_{q_3}$ is related to ${\cal K}_{4}$ therein. We have not discovered any new independent non-trivial symmetry operators of this type in the vacuum case.}

\subsection{Tensor symmetry operators: summary}
In addition to the above operators, we also have the cubic in $k$ secondary Killing vector operator:  
\be
O^{(t)}_{\zeta}=
 \zeta^e\nabla_e \delta^c_a \delta^d_b-
(d\zeta)^c{}_a \delta^d_b=\delta^c_a \delta^d_b {\cal L}_{\zeta}=\delta^c_a \delta^d_b\partial_{\varphi}\,.
\ee
Of course, such an operator is valid both on-shell and off-shell. 
While we have not checked this explicitly, we expect that, similar to the vector case, this is the only cubic in $k$ symmetry operator.

If so, this means that, we have found the following {7 operators:
\be
\{O^{(t)}_{\xi}\,,\ O^{(t)}_{l_1}\,,\  O^{(t)}_{l_2}\,,\  O^{(t)}_{q_1}\,,\ O^{(t)}_{q_2}\,,
O^{(t)}_{q_2}, O^{(t)}_{\zeta}\}\,,
\ee
that commute with the Lichnerowicz operator for the on-shell Kerr--NUT-AdS spacetimes with arbitrary cosmological constant $\Lambda$. Together with $L$, we thus have a set of 8 operators.
It remains to be seen, if these are enough to directly separate gravitational perturbations in these spacetimes.   }

Interestingly, for the off-shell Kerr--NUT--AdS spacetimes, only the 2 Killing vector operators commute with the Lichnerowicz operator. Perhaps this could be amended by adding curvature terms into our collection of potential operators -- the question boils down to whether these terms are independent. Since we have considered both orderings of the covariant derivatives this will in principle generate curvature terms for an arbitrary linear combination of our operators. Moreover, although the commutator of covariant derivatives acting on a two tensor has two independent contractions of the Riemann tensor, a closed conformal Killing--Yano tensor satisfies various integrability properties (see Appendix C.2 of \cite{Frolov:2017kze}) which may mean this terms are not independent.
We leave this for future study.

\section{Summary}\label{Sec:summary}
In this work, we have systematically studied homogeneous symmetry operators for scalar, vector, and tensor perturbations of (general) four-dimensional  {Kerr--NUT--AdS spacetimes, which are known to admit a fundamental hidden symmetry of the principal Killing--Yano tensor $k$.} In particular, we have concentrated on operators of degree two in the sum of the number of derivatives acting on $k$ and acting on the corresponding perturbation. In this way we have obtained linear, quadratic, and cubic in $k$ operators that commute with the corresponding field operator for both off-shell and on-shell Kerr--NUT--AdS metrics. 

{
Our study has developed upon the recent work \cite{Mei:2023pho} -- we have generalized it  in a number ways.
First, the work in \cite{Mei:2023pho} fully concentrated on the vacuum Kerr case, whereas ours deals with more general (possibly off-shell) Kerr--NUT--AdS spacetimes. Second, we have found a number of new (predominantly linear)  symmetry operators. 
Third, we have directly worked with the principal tensor $k$ (rather than its Hodge dual $*k$) which is the fundamental  symmetry of the Kerr--NUT--AdS spacetimes that survives in any number of dimensions. 

For this reason, our results readily generalize to higher dimensions -- the operators will take the same form as those in 4d, apart from dimension-dependent factors. For example, the new linear in $k$ 4d operator \eqref{newlinear} generalizes to 5d off-shell Kerr--NUT--AdS spacetimes as follows:  
\be\label{newlinear2}
O^{(v)}_k= k^a{}_b \nabla^2-k_{cb}\nabla^c \nabla^a
-k^{ac}\nabla_c\nabla_b-\frac{9}{4}\xi_b \nabla^a\,,
\ee
that is, the last factor simply changed from $(-2)$ to $(-9/4)$. 

In higher dimensions, however, apart from generalizing the obtained operators mentioned above that are linear, quadratic, and cubic in $k$,  one will also have to consider operators that are higher-order in $k$. For example, for the scalar case in $d$ dimensions, the complete set of commuting operators involves operators that are homogeneous in all possibles powers of $k$, ranging from zero (the box operator) to $(d-1)$. We shall return to this issue in the follow up paper \cite{myprep:2024}.   
}

{The main purpose of studying symmetry operators is to probe whether one has enough `symmetries' that would allow for separability of the corresponding test field equations. 
Contrary to for example \cite{Teukolsky:1972my,Teukolsky:1973ha}, 
in our approach we directly work with perturbations themselves (with the vector potential or the metric perturbation). In this way, we have found a set of 7 mutually commuting operators for the off-shell vector case and {a set of 8 operators for the vacuum  tensor case.}
Interestingly, for the scalar and Dirac perturbations one only needs 4 such operators  -- the separated solution is simply the common eigenfunction of such operators.  It remains to be seen, if the obtained symmetry operators `are enough' to directly separate the (massive) vector and tensor perturbations in these spacetimes.
}

\subsection*{Acknowledgements}
D.K. is grateful for support from Grant No. GA{\v C}R 23-07457S of the Czech Science Foundation.

\appendix
\section{Dirac symmetry operators}\label{App:Dirac}
In this appendix we review, following \cite{Carter:1979fe, Oota:2007vx, Cariglia:2011qb}, the symmetry operators responsible for separability of the massive Dirac equation in Kerr--NUT--AdS spacetimes. As it turns out, these are also given by the homogeneous first-order and 2nd-order operators.
For simplicity, we restrict to the four-dimensional case \cite{Carter:1979fe}, and refer to \cite{Oota:2007vx, Cariglia:2011qb} for a generalization to higher dimensions.

The Dirac operator reads 
\be
D=\gamma^{A} \nabla_{A}\,, 
\ee
where $A, B,\dots$ refer to the orthonormal frame indices, and the 
$\gamma$-matrices obey 
\be
\gamma^{A}\gamma^{B}+\gamma^{B}\gamma^{A}=2 g^{{A}{B}}\,. 
\ee
Then, in 4d off-shell canonical Kerr--NUT--AdS spacetimes, we find the following linear in PKY and linear in derivatives operator:
\be\label{Mop}
M=\gamma^{AB C}k_{{A}{B}}\nabla_{{C}}+2\xi_{A} \gamma^{A}\,, 
\ee 
where we use $\gamma^{AB\dots C}=\gamma^{[A}\gamma^{B}\dots \gamma^{C]}$ to denote antisymmetrization of the $\gamma$-matrices.
Of course, we also have the symmetry operators associated with the Killing directions. Namely, we have a linear in PKY and quadratic in derivatives operator
\be
K_\xi=\xi^{A}\nabla_{A}+\frac{1}{8}\gamma^{A B}(d\xi)_{{A}{B}}\,, 
\ee
and a cubic in PKY and quadratic in derivatives operator 
\be
K_\zeta=\zeta^{A}\nabla_{A}+\frac{1}{8}\gamma^{A B}(d\zeta)_{{A}{B}}\,.
\ee
These operators form a set of 4 mutually commuting operators:
\be\label{setDirac}
\{D,M, K_\xi,K_\zeta\}\,. 
\ee
Interestingly, such a set is enough to characterize the separability of the Dirac equation, whose separated solution can be found as a common eigenfunction of these 
operators.

In particular, introducing the following natural orthonormal basis:
\be
g=\eta_{{A}{B}}e^{A}e^{B}\,, 
\ee
where 
\ba
e^{ 0}&=& \sqrt{\frac{\Delta_r}{\Sigma}}(d\tau+y^2d\psi)\,,\quad e^{ 1}=\sqrt{\frac{\Sigma}{\Delta_r}}dr\,,\nonumber\\
e^{\hat 2}&=&
\sqrt{\frac{\Delta_y}{\Sigma}}(d\tau-r^2d\psi)\,,\quad e^{ 3}=\sqrt{\frac{\Sigma}{\Delta_y}}dy\,,
\ea
and choosing the following representation
for the $\gamma$ matrices:
\ba
\gamma^{ 0}&=&\left(   
\begin{array}{cc}
0 & -I\\
I& 0
\end{array}
\right)\, ,\quad
\gamma^{ 1}=\left(   
\begin{array}{cc}
\ 0 & \ I\\
\ I& \ 0
\end{array}
\right)\,, \nonumber\\
\gamma^{ 2}&=&\left(   
\begin{array}{cc}
\sigma^2 & 0\\
0& -\sigma^2
\end{array}
\right)\, ,\quad
\gamma^{ 3}=\left(   
\begin{array}{cc}
\sigma^1 & 0\\
0& -\sigma^1
\end{array}
\right)\, ,
\ea
where $I$ is a unit $2\times 2$ matrix and $\sigma^i$ are Pauli matrices,
the {\em R-separated}~\footnote{R-separability is a generalized notion of multiplicative separability, where each component of the field takes a multiplicative separable form, up to a `known'  non-separable factor (the so called $R$-factor), which is in principle a function of all coordinates, and has to be `guessed', see \eqref{psiAnsatz} above. } solution $\Psi$ of the Dirac massive equation:
\be
(D+\mu)\Psi=0\,, 
\ee
is the eigenfunction of the following eigenvalue problem: 
\ba
D\Psi &=&-\mu\Psi\,,\nonumber\\
M\Psi&=&\kappa\Psi\,,\nonumber\\
K_\xi\Psi&=&-i\omega \Psi\,,\nonumber\\
K_\zeta\Psi&=&im\Psi\,,
\ea
where $\omega, m$ stands for the frequency, azimuthal number, respectively, and $\kappa$ is the separation constant. Namely, it reads 
\be\label{psiAnsatz}
\Psi=\left(   
\begin{array}{c}
(r-iy)^{-1/2}R_+Y_+\\
(r+iy)^{-1/2}R_+Y_-\\
(r+iy)^{-1/2}R_-Y_+\\
(r-iy)^{-1/2}R_-Y_-
\end{array}
\right)\,e^{i(m\psi-\omega t)}\,,
\ee
with functions $R_{\pm}=R_{\pm}(r)$ and $Y_{\pm}=Y_{\pm}(y)$ obeying the following ODEs, e.g. \cite{Frolov:2017kze}: 
\ba
\frac{d R_\pm}{dr}+R_\pm\frac{\Delta_r'\pm V_r}{4\Delta_r}+R_{\mp}\frac{\mu r\mp \kappa}{\sqrt{\Delta_r}}&=&0\,,\nonumber\\
\frac{d Y_\pm}{dy}+Y_\pm\frac{\Delta_y'\pm V_y}{4\Delta_y}-Y_{\mp}\frac{\kappa\pm i\mu y}{\sqrt{\Delta_y}}&=&0\,,
\ea
where $V_r$ and $V_y$ are given by 
\be
V_r = 4i(m-\omega r^2)\,,\quad V_y=4(m+\omega y^2)\,. \label{Upot}
\ee

Let us stress that in principle the ansatz \eqref{psiAnsatz} leads to 8 equations with four different separation constants. However, the consistency of these equations implies that only 1 of these separation constants is independent, given by the eigenvalue of the operator $M$. In other words, with the special separability ansatz above, the 4 operators \eqref{setDirac} form a complete set of commuting operators responsible for separation of the Dirac equation. A similar situation happens in higher dimensions, where a set of $D$ mutually commuting operators underlies separability of the Dirac equation in $D$-dimensional Kerr--NUT--AdS spacetimes \cite{Oota:2007vx, Cariglia:2011qb}.  
This raises the hope that perhaps a similar statement is, with a proper separation ansatz, also true for other (vector or tensor) fields.

\section{Motion of particles}\label{App:motion}
The pattern seen for operators of various fields also translates to quantities associated with motion of particles:  homogeneous operators translate to  corresponding  integrals of motion. Let us review this for geodesics and for the motion of a classical spinning particle.

\subsection{Geodesics}
The geodesic motion is described by the following Hamiltonian: 
\be
H=\frac{1}{2}g^{ab}p_ap_b\,, 
\ee
where $p_a(\tau)$ are canonical momenta, conjugate to particle's coordinates $x^a(\tau)$. The Hamilton's equations of motion are equivalent to the geodesic equation.

To find an integral of motion $Q$, we seek a quantity that obeys 
\be\label{QH}
\{Q, H\}=0\,,
\ee
where the Poisson brackets are defined as 
\be
\{F,G\}=\frac{\partial F}{\partial x^a}\frac{\partial G}{\partial p_a}-
\frac{\partial G}{\partial x^a}\frac{\partial F}{\partial p_a}\,. 
\ee
We may proceed in a way similar to what happens in the main text, seeking a scalar quantity that is homogeneous in the number of PKY tensors and homogeneous in the `sum of' derivatives acting on $k$ and the number of momenta $p_a$.

It is easy to see that there are no quantities that are linear in $k$ and {of degree one.} There is one quantity that is linear in $k$ and {of degree two:} 
\be
Q_\xi=\xi^a p_a\,,
\ee
which corresponds to a constant of motion associated with the Killing vector $\xi$. 
Moving to quadratic in $k$, there are no degree one quantities, but there are the following degree two quantities:
\be 
(k^2)^{ab}p_a p_b\,,\quad k^2p^2\,,\quad 
k^a{}_b \xi^b p_a\,,\quad 
\xi^2\,,\quad (d\xi)^{ab}k_{ab}\,,
\ee 
By requiring \eqref{QH}, the last 3 have to be eliminated, and we recover the famous Carter's constant \cite{Carter:1968rr}
\be
Q_K=\Bigl((k^2)^{ab}+\frac{1}{2}k^2g^{ab}\Bigr)p_a p_b=K^{ab}p_ap_b\,. 
\ee

Considering finally constants that are cubic in $k$ and of degree two, we have the following non-trivial possibilities:
\be
(k^2)^a{}_b \xi^b p_a\,,\quad k^2\xi^ap_a\,.
\ee
These combine to the following constant of motion:
\be
Q_\zeta=
\Bigl((k^2)^a{}_b \xi^b +\frac{1}{2}k^2\xi^a\Bigr)p_a=\zeta^a p_a \,,
\ee
associated with the Killing vector $\zeta$, 
which finalizes the complete set of mutually commuting integrals of motion:
\be 
\{H, Q_\xi, Q_\zeta, Q_K\}\,,
\ee
which underlie the complete integrability of geodesic motion in Kerr--NUT--AdS spacetimes.

\subsection{Classical spinning particle}
The above construction for geodesics can be generalized to classical spinning particles, e.g. \cite{Rietdijk:1989qa,Gibbons:1993ap}. Such a motion can be derived from the following Lagrangian:
\be
{\cal L}=\frac{1}{2}g_{ab}\dot x^a\dot x^b + \frac{i}{2}\eta_{A B}\theta^A\frac{D\theta^B}{d\tau}\,,
\ee
where $x^a(\tau)$ ($a=1,\dots , d$)
denote the particle's worldline coordinates,  and the spin is described by a Lorentz vector of Grassmann-odd coordinates  $\theta^A(\tau)$ ($A = 1, \dots, d$). In this subsection, $d$ stands for the spacetime dimension, indices $A,B,\dots$ denote vielbein indices, indices $a,b,\dots$ denote the curved space indices, and $e^a_A$ are  the veilbein components (that can be used to convert spacetime to vielbein indices and vice versa). 
The above Lagrangian yields the following equations of motion:
\ba
\frac{D^2x^a}{d\tau^2}&=&\ddot{x}^a + \Gamma^a_{bc} \dot{x}^b \dot{x}^c = \frac{i}{2} R^a_{\ b A B}\theta^{A}\theta^B\dot x^b\,,
\label{eq:Papapetrou}\\
\frac{D \theta^A}{D\tau} &=& \dot{\theta}^A + \omega_b{}^A{}_B \dot{x}^b \theta^B = 0 \, , \label{eq:covariantly_constant_spin}
\ea
where $\Gamma^a_{bc}$ and $\omega_b{}_{AB}$ are the Levi-Civita and spin connections, respectively, and 
$R_{abcd}$ is the Riemann tensor.

The theory possesses a generic {\em supercharge} $Q$,
\be\label{Qdef}
Q=\theta^A e_A{}^a \Pi_a\,,
\ee
which obeys
\be
\{H,Q\}=0\,, \quad \{Q,Q\}=-2iH\,.
\ee
Here, $H$ is the Hamiltonian,
\be
H=\frac{1}{2}\Pi_a\Pi_b g^{ab}\,,
\ee
where
$\Pi_a$ is 
related to $p_a$, the momentum canonically conjugate to $x^a$, as follows:
\be
\Pi_a=p_a - \frac{i}{2}\theta^A\theta^B\omega_{a AB}=g_{ab}\dot x^b\,, 
\ee
and the Poisson brackets are defined as
\be\label{brackets}
\{F,G\}=\frac{\partial F}{\partial x^a}\frac{\partial G}{\partial p_a}-
\frac{\partial F}{\partial p_a}\frac{\partial G}{\partial x^a}+
i(-1)^{a_F}\frac{\partial F}{\partial \theta^A}\frac{\partial G}{\partial \theta_A}\,,
\ee
where $a_F$ is the Grassmann parity of $F$,
and the equations of motion are accompanied by two
 physical (gauge) conditions
\be\label{gaugecond}
2H=-1\,,\quad Q=0\,,
\ee
which state that $\tau$ is the proper time and the particle's spin is spacelike.

A non-generic superinvariant $S$ is a quantity that Poisson commutes with the generic supercharge: 
\be\label{Q}
\{Q, S\}=0\,.
\ee
Due to the Jacobi identity, any superinvariant is automatically a constant of motion,  $\{H, S\}=0$, and so is a new superinvariant $\{S,S\}$ (which may, or may not be equal to $H$). Such superinvariants 
correspond to an enhanced worldline (super)symmetry.

As shown in \cite{Gibbons:1993ap, Kubiznak:2011ay}, for Kerr--NUT--AdS spacetimes, the spinning particle motion admits a number of homogeneous superinvariants. In particular, focusing on 4d, we have the following two superinvariants linear in the PKY tensor:
\ba
Q_\xi&=&\xi^a \Pi_{a}-\frac{i}{4}\theta^A \theta^B(d\xi)_{A B}\,,\\ 
Q_k&=&\theta^A(*k)^a{}_A\Pi_a-\frac{i}{9}\theta^A\theta^B\theta^C (d*k)_{A B C}\,.
\ea 
We also have a quadratic in $k$ superinvariant
\be
Q_{k^2}=K^{ab}\Pi_a\Pi_b+\theta^A\theta^B L_{AB}{}^a \Pi_a+\theta^A\theta^B\theta^C\theta^D M_{ABCD}\,, 
\ee
where 
\ba
K^{ab}&=&(*k)^{ae}(*k)^b{}_e
=k_{ac} k_b^{\ c}+\frac12 k^2 g_{ab}\,,
\nonumber\\
L_{abc}&=&-\frac{2i}{3}(*k)_{[a|e}(d*k)_{b]c}{}^{e}-\frac{2i}{3}(d*k)_{abe}(*k)_{c}{}^{e}\,,
\nonumber\\
&=&2i \big( \xi_{[a} k_{b]}{}_{c} +k_{a b} \xi_c+ (\xi\cdot k)_{[a}g_{b]c}
 \big)
\nonumber\\
M_{abcd}&=&-\frac{i}{4}\nabla_{[a}L_{bcd]}=-\frac i 2 d\xi_{[a b}k_{c d]}\,.
\ea
It also admits a cubic in $k$ superinvariant  
associated with the Killing vector $\zeta$:
\be
Q_\zeta=\zeta^a \Pi_{a}-\frac{i}{4}\theta^A \theta^B(d\zeta)_{AB}\,. 
\ee
Note that $\{Q_\xi, Q_{k^2}, Q_\zeta\}$ are Grassmann even (bosonic) and generalize the above constants of geodesic motion. One can show that in higher dimensions, there will always be $d$ such bosonic constants for the $d$-dimensional spinning particle motion, see  \cite{Kubiznak:2011ay}. 
On the other hand superinvariants $\{Q, Q_h\}$ are Grassmann odd (fermionic), and are reminiscent of the Dirac operator $D$ and the linear in $k$ operator $M$, \eqref{Mop}.

\section{Working with the Hodge dual?}\label{App:f}
{
In the main text, we have fully focused on working with the principal tensor $k$. This is primarily motivated by the fact that $k$ is the fundamental symmetry of the Kerr--NUT--AdS spacetimes that has the same rank (of a 2-form) in any number of dimensions, whereas this is not true for its (dimension-dependent) Hodge dual. However, one might wonder whether working with $(*k)$ would enlarge the number of commuting operators.\footnote{Of course, this may not be the case for quadratic (or even power) in $(*k)$ operators, as in that case, one can always eliminate the Hodge star and work directly with $k$'s.
}
In this appendix, we gather a simple calculation for the linear in $(*k)$ vector operators in 4d, showing that no new symmetry operators arise in this case.   
}

{
When working with the Hodge dual of $k$, 
\be
f=*k 
\ee
we have the following objects at our disposal:
\be
f_{ab}\,,\quad (df)_{abc}\,,\quad 
\nu^{ab}=\nabla_e (df)^{abe}\,,
\ee 
capturing the full information about the zeroth-order, 1st-order, and 2nd-order derivatives of $f$. For the linear in $f$ degree two operators we thus have the following 8 possibilities: 
\ba
O_1&=&f^a{}_b \nabla^2\,,\quad 
O_2= f_{cb}\nabla^a \nabla^c\,,\quad 
O_3= f_{cb}\nabla^c \nabla^a\,,\nonumber\\
O_4&=& f_{cd}\nabla^c \nabla^d \delta^a_b\,,\quad 
O_5=f^{ac}\nabla_c\nabla_b\,,\quad 
O_6=f^{ac}\nabla_b\nabla_c\,,
 \nonumber\\
O_7&=&(df)^a{}_{be}\nabla^e\,,\quad 
O_8=\mu^a{}_b\,.
\ea 

One can then easily show that requiring the commutativity of the vector operator with the linear combination of these operators, \eqref{Vcomm}, does not yield any non-trivial commuting operators in the off-shell Kerr--NUT--AdS spacetimes. Thus, using $f$ in addition to $k$ does not bring any new operators in the linear vector case. We suspect (but have not checked explicitly), that the same conclusion remains true also in other (for example tensor) cases.  
}

\bibliography{Databaze}
\bibliographystyle{JHEP}

\end{document}